\documentclass[aps,prd,amsmath,amssymb,twocolumn,preprintnumbers,nofootinbib]{revtex4-1}
\pdfoutput=1
\usepackage{graphicx}
\usepackage{amsthm}
\usepackage{amsmath}
\usepackage{graphicx,subfigure}
\usepackage{dcolumn}
\usepackage{bm}
\usepackage{slashed}
\usepackage{enumitem}

\def\ba{\begin{eqnarray}}
\def\ea{\end{eqnarray}}
\def\be{\begin{equation}}
\def\ee{\end{equation}}
\def\nn{\nonumber}
\def\exd{{\rm d}}
\def\pd{\partial}
\makeatletter
\def\x@arrow{\DOTSB\Relbar}
\def\xlongequalsignfill@{\arrowfill@\x@arrow\Relbar\x@arrow}
\newcommand{\xlongequal}[2]{%
    \ext@arrow 0099\xlongequalsignfill@{#1}{#2}}
\makeatother

\newcommand{\roughly}[1]{\mathrel{\raise.3ex\hbox{$#1$\kern-0.85em
\lower1ex\hbox{$\sim$}}}}

\newcommand{\lsim}{\roughly<}
\newcommand{\gsim}{\roughly>}

\def\nott#1{\setbox0=\hbox{$#1$}                
   \dimen0=\wd0                                 
   \setbox1=\hbox{/} \dimen1=\wd1               
   \ifdim\dimen0>\dimen1                        
      \rlap{\hbox to \dimen0{\hfil/\hfil}}      
      #1                                        
   \else                                        
      \rlap{\hbox to \dimen1{\hfil$#1$\hfil}}   
      /                                         
   \fi}                                         %

\def\endignore{}
\def\ignore #1\endignore{} 

\def\be{\begin{equation}}
\def\beq\begin{equation}
\def\ee{\end{equation}}
\def\bea{\begin{eqnarray}}
\def\eea{\end{eqnarray}}

\def\eqa{\begin{eqnarray}}
\def\eeqa{\end{eqnarray}}
\def\eq{\begin{equation}}
\def\eeq{\end{equation}}

\def\nn{\nonumber}

\def\pref#1{(\ref{#1})}

\def\exd{{\rm d}}
\def\nn{\nonumber}
\def\pref#1{(\ref{#1})}
\def\be{\begin{equation}}
\def\ee{\end{equation}}

\def\beq{\begin{equation}}
\def\eeq{\end{equation}}
\def\beqa{\begin{eqnarray}}
\def\eeqa{\end{eqnarray}}

\def\cA{{\cal A}}

\def\cC{{\cal C}}

\def\cL{{\cal L}}

\def\cO{{\cal O}}

\def\cV{{\cal V}}

\def\ssA{{\scriptscriptstyle A}}
\def\ssB{{\scriptscriptstyle B}}

\def\ssE{{\scriptscriptstyle E}}
\def\ssF{{\scriptscriptstyle F}}

\def\ssL{{\scriptscriptstyle L}}
\def\ssM{{\scriptscriptstyle M}}
\def\ssN{{\scriptscriptstyle N}}

\def\ssR{{\scriptscriptstyle R}}

\def\ssV{{\scriptscriptstyle V}}

\def\SM{{\scriptscriptstyle SM}}

\newcommand{\bmat}{\left(\begin{array}}
\newcommand{\emat}{\end{array}\right)}

\def\-{\hphantom{-}}

\def\s2{\frac{1}{2}}

\def\IF{\relax{\rm I\kern-.18em F}}
\def\II{\relax{\rm I\kern-.18em I}}
\def\IP{\relax{\rm I\kern-.18em P}}
\def\IC{\relax{\rm I\kern-.48em C}}
\def\IR{\relax{\rm I\kern-.18em R}}
\def\IK{\relax{\rm I\kern-.20em K}}
\def\IM{\relax{\rm I\kern-.25em M}}

\def\y2{Y_{\ssM\ssN} Y^{\ssM\ssN}}
\def\Riem2{R_{\ssA\ssB\ssM\ssN} R^{\ssA\ssB\ssM\ssN}}
\def\Ricci2{R_{\ssM\ssN} R^{\ssM\ssN}}

\def\f2{F^{a}_{\ssM\ssN} F^{\ssM\ssN}_a}

\def\Asl{\hbox{/\kern-.7500em\it A}} 
\def\dsl{\hbox{/\kern-.5500em$\partial$}}
\def\pxpsl{\hbox{/\kern-.5600em$p$}}
\def\Dsl{\,\raise.15ex\hbox{/}\mkern-13.5mu D}
\def \one{\relax{\rm 1\kern-.26em I}}

\def\exd{{\rm d}}

\def\nn{\nonumber}

\def\({\left(}
\def\){\right)}

\def\Vone{\cV_{1\ssL}}
\def\gR{g_\ssR}

\begin{document}

\title{Technically Natural Vacuum Energy at the Tip of a Supersymmetric Teardrop}
\author{Matthew Williams}
\email{mwilliams@perimeterinstitute.ca}
\affiliation{Department of Physics \& Astronomy, McMaster University \\ 1280 Main St.~W., Hamilton, Ontario, Canada, L8S 4L8}
\affiliation{\quad\quad\quad\quad Perimeter Institute for Theoretical Physics\quad\quad\quad\quad \\ 31 Caroline St.~N., Waterloo, Ontario, Canada N2L 2Y5.}

\date{\today}

\begin{abstract}

A minimal supersymmetric brane-world model is presented which has: i) zero classical four-dimensional vacuum curvature, despite the large na\"ive vacuum energy due to contributions from Standard Model particles; ii) one-(bulk)-loop quantum corrections to the vacuum energy with a size set by the radius of the extra-dimensional spheroid. These corrections are technically natural because a BPS-like relation between the brane tension and $R$-charge---which would have preserved (half of) the bulk supersymmetry---is violated by the requirement that the stabilizing $R$-symmetry gauge flux be quantized. The extra-dimensional geometry is similar to previous rugby-ball geometries, but is simpler in that there is only one brane and so fewer free parameters. Although the sign of the renormalized vacuum energy ends up being the unphysical one for this model (in the limit considered here, where the massive bulk loop is the leading contribution), it serves as an illustrative example of the relevant physics.

\end{abstract}

\maketitle


\section{Introduction}

One of the many reasons why a technically-natural solution to the cosmological constant problem has eluded physicists is because of the difficulty in achieving a type of `Goldilocks' criterion: a physical theory which predicts a value for the vacuum energy ought to give one that is not simply zero, but still many orders of magnitude below the physical scales relevant to particle physics. (For a pedagogical review of the relevant issues, see e.g.~\cite{CliffsDENotes}.) There are very few theories which predict a value for the vacuum energy that is `just right'.

The reason for this is that it is very difficult to disentangle the energy density of a Lorentz-invariant vacuum from its corresponding curvature, as dictated by Einstein's equation. A possible loophole is to consider a construction similar to that of a cosmic string \cite{Vil}, wherein energy localized on the string only curves the dimensions transverse to string. In analogy, if there were two  dimensions transverse to our familiar four dimensions, a large vacuum energy from loops of Standard Model particles confined to our space would curve these extra dimensions and not our own \cite{preSLED}. However, this cannot be the full story since massive bulk loops generally yield unacceptably large corrections to the four-dimensional curvature in the absence of some symmetry, such as supersymmetry, to forbid or suppress such corrections \cite{Towards}.

In \cite{TechNat}, it was shown that---in a particular six-dimensional brane-world supergravity model---zero curvature can be obtained at the classical level, despite large contributions to the na\"ive vacuum energy from Standard Model particles. This result is guaranteed to persist as one integrates out Standard Model particles by simply requiring that the brane lagrangian does not couple to the scalar in the bulk gravity multiplet, commonly referred to as the dilaton (see also \cite{OtherConical}). Subsequently in \cite{Companion}, quantum corrections due to bulk physics are shown under certain conditions to contribute a positive vacuum energy whose order of magnitude is set by the size of the extra dimensions. 

This model's success relies heavily on three main ingredients: 1) a stabilizing flux in the extra dimensions, which couples to the branes; 2) a supersymmetric extra-dimensional sector (which includes gravity); and 3) a careful treatment of back-reaction when considering brane physics. There are, however, some features---such as the specific shape of the extra dimensions---which do not play such an important role. In fact, in the case of the rugby-ball-shaped extra dimensions employed therein, a second brane is a hindrance. A non-zero vacuum energy is obtained only if the two branes (with identical tensions) have different charges. Therefore, the presence of an additional, unspecified parameter (namely, the difference of the two brane charges) muddles the relationship between the size of the extra dimensions and the observed vacuum energy. The teardrop model considered here---which is similar to non-supersymmetric \cite{GellMann} and supersymmetric \cite{KehagiasTD} versions proposed previously---uses these aforementioned desirable ingredients to obtain a more direct relationship between these quantities. 

The purpose of this paper is twofold: i) to lay out the properties of this simple model so that it can serve as a point of comparison for future work, such as the development of a four-dimensional effective theory of supersymmetric large extra dimensions \cite{4DSLED}; ii) to demonstrate the applicability of the results of \cite{Companion} in more general contexts (see also \cite{LittlestGravitino}); iii) to estimate the contribution due to a massive bulk loop in the limit where its Compton wavelength is much smaller than the size of the extra dimensions. Regarding the latter, the analysis presented herein makes it seems rather unavoidable that such a contribution will yield a negative vacuum energy; as such, it remains a pedagogically useful, albeit unrealistic extra-dimensional theory. (There exist limits---such as when the mass of the multiplet is sent to zero---where the analysis presented here does not apply; further consideration is needed to draw conclusions in those cases.) Despite this shortcoming, an order-of-magnitude analysis shows that the observed magnitude of the vacuum energy can be achieved with a gravity scale $M_g\gsim 14 \,{\rm TeV}$, which corresponds to extra dimensions with size $r\lsim 0.7 \,\mu{\rm m}$. These estimates evade the known astrophysical ($M_g \gsim 10$ TeV) \cite{Raffelt} and inverse-square-law ($r \lsim 45 \mu$m) \cite{Adelberger} bounds on large extra dimensions. 

In Section \ref{sec:SSS}, we briefly review the Salam-Sezgin solution \cite{SS}. In Section \ref{sec:Teardrop}, we describe the effects of matching a single codimension-2 brane to the background in the bulk. Section \ref{sec:ModeSums} reviews the procedure for computing the effects of bulk loops on the four-dimensional vacuum energy, and Section \ref{sec:MassMult} computes the result of a massive multiplet loop in the case of the teardrop. Section \ref{sec:Conclusion} provides a summary of these results, and proposes directions for further research.

\section{Salam-Sezgin Solution}
\label{sec:SSS}

In the Salam-Sezgin solution \cite{SS} to 6D chiral, gauged supergravity, the non-trivial background fields are the graviton $g_{\ssM\ssN}$, the $R$-symmetry gauge field $F_{\ssM\ssN}$, and the dilaton $\phi$. Their equations of motion follow from the action\footnote{The conventions are: natural units ($\hbar=c=1$), a `mostly-plus' metric, the Weinberg curvature conventions \cite{GandC}, and indices where $M,N,\ldots \in \{t,x,y,z,\theta,\varphi\}$; $\mu,\nu,\ldots \in \{t,x,y,z\}$; $m,n,\ldots \in \{\theta,\varphi\}$.}
\bea
S &=& -\int \! \exd^6 X \sqrt{-g} \bigg[ \frac1{2\kappa^2} \Big( R + \pd_\ssM \phi \, \pd^\ssM\phi \Big) \nn\\
&&\qquad\qquad\qquad+ \frac{e^{-\phi}}{4 \gR^2} \, F_{\ssM\ssN} \, F^{\ssM\ssN} + \frac{2 \gR^2}{\kappa^4} \, e^\phi \bigg] \,,
\eea
and---along with the \emph{ans\"atze}
\begin{align}
\exd s^2 &= \hat g_{\mu\nu} \, \exd x^\mu \exd x^\nu + r^2 (\exd \theta^2 + \sin^2\theta \, \exd\varphi^2 ) \\
F_{\mu\ssN} &= 0 \,,\quad F_{mn} = f \, \epsilon_{mn} \,,\quad \phi = {\rm const.} \,,
\end{align}
where $r$ is the radius of the extra dimensions, $\hat g_{\mu\nu}$ is maximally-symmetric, and $\epsilon_{mn}$ is the volume form of the $2$-sphere---result in the following solution:
\be
\hat g_{\mu\nu} = \eta_{\mu\nu} \,,\;\; F_{\theta\varphi} = \pm \frac{\epsilon_{\theta\varphi}}{2r^2} = \pm \frac{\sin\theta}2 \,,\;\; r^2 e^\phi = \frac{\kappa^2}{4 \gR^2} \,.
\ee
This is a 1-parameter family of solutions for some given $\kappa$ and $\gR$; it is possible to obtain different solutions by scaling $r$ and shifting $\phi$:
\be
r \to  k r \,,\quad \phi \to \phi -2\ln k \,.
\ee
This freedom can be removed by introducing a four-dimensional brane, as we shall see next.

\section{The Teardrop}
\label{sec:Teardrop}

Introducing a four-dimensional source in this space will generally distort the spherical Salam-Sezgin solution. We model such a source---inserted at $\cos\theta=+1$---with the following effective brane action\footnote{We initially use notation similar to ref.~\cite{Companion}; however, since there is no brane at $\cos\theta=-1$, $T_- = \cA_- = 0$ herein so we shall drop the `$+$' subscripts later on.}:
\be
S_+ = \int \exd^4x \sqrt{-g_4} \left(-T_+ + \frac{\cA_+}{2\gR^2} \, \epsilon^{mn} F_{mn} \right) \,.
\ee
The first term is the usual tension, which---in this model---would get contributions from any Standard Model particle of size $T \sim m_\SM^4$; the second term represents a charge carried by the four-dimensional source, which is distinct from those of the Standard Model fields, and which takes a form analogous to a Wilson line. (The latter can be written using differential forms as $\tfrac{\cA_+}{2\gR^2} \!\int {}^* F$, where ${}^*$ denotes the Hodge dual.)

In the presence of this source, the near-brane metric develops a conical singularity:
\be
\exd s^2_{\theta \sim 0} = W_+^2 \,\eta_{\mu\nu} \, \exd x^\mu \exd x^\nu + r_+^2 (\exd \theta^2 + \alpha^2 \theta^2 \,\exd \varphi^2 ) \,.
\ee
(In the above, $W_+$ is the warp factor evaluated at the brane; for details about metric at macroscopic distances from the brane, see the appendix.) The corresponding deficit angle, which describes the size of the angular section removed at the singularity, is given by
\be
\delta = 2\pi (1-\alpha) \,.
\ee

By matching across a codimension-1 surface surrounding the source \cite{localizedflux}, we find that---in a gauge where the polar component of $A_m$ is set to zero---the gauge field boundary condition is\footnote{Another notational comment: we keep the `$+$' subscript on $\phi$ here because we wish to distinguish its value at $\cos\theta=+1$ from its value at other points in the extra dimensions; for more information, see the appendix.}
\be \label{Phidef}
A_\varphi \Big|_{\cos\theta=+1} = \frac{\cA \,e^{\phi_+}}{2\pi} := \Phi \,.
\ee
Similarly, $\alpha$ is given by 
\be
1-\alpha = \frac{\kappa^2 T}{2\pi} \mp 2 \Phi \,.
\ee
However, the flux quantization condition---which was already enforced by the equations of motion  in the Salam-Sezgin case, since $\int F = \pm 2\pi$ there---now becomes\footnote{In principle, the r.h.s.~need only be an integer, not necessarily $\pm 1$. However, in order to stay perturbatively close to the Salam-Sezgin solution, we will take its value to be unchanged in the case of the teardrop.}
\be
\Phi + \frac1{2\pi}\!\int F  =\pm 1 \,.
\ee
Since the volume is deformed by having $\alpha\neq 1$, this condition is not inconsistent, but instead determines $\Phi$:
\bea
\Phi &=& \pm \left(1-\sqrt{1-\frac{\kappa^2 T}{2\pi} \pm 2\Phi}\right)  \nn\\
\implies \Phi &=& \pm 2 \left(1-\sqrt{1-\frac{\kappa^2 T}{8\pi}}\right) \,.
\eea
(The integral over $F$, in the case of a general warped geometry, is proportional to $\alpha_{\rm gen} = \sqrt{\alpha_+\alpha_-}$ \cite{TechNat}; however, $\alpha_-$ is unity in the case of the teardrop. See the appendix for details.) Thus, the dilaton field $\phi$ adjusts to satisfy the above for any $T$ and $\cA$:
\bea
e^{\phi_+} &=& \frac{4\pi\bigg(1-\sqrt{1-\frac{\kappa^2 T}{8\pi}}\bigg)}{|\cA|} \nn\\
&\simeq&  \frac{\kappa^2 T}{4|\cA|} + \frac{(\kappa^2 T)^2}{128\pi|\cA|} + \cdots  \label{ephipval}
\eea
In other words, the residual shift symmetry in $\phi$ is fixed so that flux quantization is satisfied for any non-zero values of the tensions and charges.

Some supersymmetry at the brane would have remained unbroken if, instead, the flux and tension were related by
\bea
\Phi_{\rm s} &=& \pm \frac12(1-\alpha) \quad\leftrightarrow\quad e^{\phi_+} = \frac{\kappa^2 T}{4|\cA|} \,, \label{ephisval}
\eea
so we see that---although the leading terms agree---flux quantization enforces the breakdown of supersymmetry. (We find later on that this agreement at leading order corresponds to more suppression than na\"ively expected.)

Given this background, we wish to compute vacuum loops of various fields comprising a massive multiplet. To this end, we review the approach used in \cite{Companion}.

\section{Mode sums and renormalization}
\label{sec:ModeSums}

Our goal is to compute the UV-sensitive part of the 1PI quantum action, $\Gamma = S + \Sigma$, due to bulk loops. Our starting point is the following expression:
\bea
    i\Sigma &=& -i \int \exd^4 x \, \Vone \nn\\
     &=& -\, \frac{1}{2} \, (-)^\ssF \, \hbox{Tr}\;
    \hbox{Log} \,\left( \frac{ -\Box_6 + X + m^2}{\mu^2} \right) \,,
    \label{eqn: sigma} 
\eea
where $m$ is the mass of the particle, $X$ denotes additional operators specific to the type of field in the loop, and where bosons/fermions contribute with $(-)^\ssF = \pm 1$;

\subsection*{One-loop mode sums}

Wick rotating to Euclidean signature and performing a heat-kernel expansion \cite{GilkeydeWitt, GdWrev}, we have
\eqa
    \Vone &=& \frac12 \, (-)^\ssF \, \mu^{4-d} \sum_{jn} \int \frac{\exd^d k_\ssE}{(2\pi)^d} \, \ln \left( \frac{k_\ssE^2 + m^2 + m_{jn}^2}{\mu^2}  \right) \nn\\
    &=& -\frac{\mu^{4-d}}{2(4 \pi r^2)^{d/2}}  \int_0^\infty \frac{\exd t}{t^{1 + d/2}} \, e^{- t (m r)^2} \, S(t) \,,
\eea
where $m_{jn}^2 = \lambda_{jn}/r^2$ denote the eigenvalues of $-\Box_2 + X$ in the compactified space and $d = 4 - 2 \, \varepsilon$ with regularization parameter, $\varepsilon$, which we eventually take to zero after all divergences in this limit are renormalized. The function $S(t)$ is defined by
\bea
 S(t) &:=& (-)^\ssF \sum_{jn} \exp \left[ - t \lambda_{jn}  \right]
\eea
and has the following small-$t$ expansion:
\bea
S(t) &\simeq&  \frac{s_{-1}}{t} +
 \frac{s_{-1/2}}{\sqrt t} + s_0 + s_{1/2}\, \sqrt{t} \nn\\
 &&\quad  + s_1 \, t + s_{3/2}\, t^{3/2} + s_2 \, t^2 + \cO(t^{5/2})\,.\quad
\eea
Its small-$t$ limit is of interest because it is only a few of the first terms in this series that contribute to the UV divergences appearing in $\Vone$:
\be \label{eq:Vinfty}
 \Vone = \frac{\cC}{(4\pi r^2)^2} \left[ \frac1{4-d} + \ln\left(\frac\mu{m}\right)\right] + \cV_f \,,
\ee
where $\cV_f$ is finite as $d \to 4$. The constant $\cC$ is given in terms of the $s_i$ by
\be \label{eq:Cform}
 \cC := \frac{s_{-1}}{6} (m r)^6 - \frac{s_0}{2}(m r)^4 + s_1 (mr)^2 - s_2 \,.
\ee
The coefficients $s_i$ are functions of the teardrop's defect angle, $\delta = 2\pi(1 - \alpha)$, and its brane charge, $\Phi = \cA \,e^{\phi_+}\!/(2\pi)$. 

These ultraviolet divergences also track the dominant dependence on $m$ in the limit that $m \gg 1/r$, since both UV divergences and large masses involve the short-wavelength part of a loop that can be captured as the renormalization of some local effective interaction. This observation will be relevant when considering the size of the final, $m$-dependent result, as it is the dominant one only in this limit. We point this out because the undetermined part, $\cV_f$, may otherwise contribute significantly to the result. We drop it in what follows, as there is evidence from similar calculations (see, {\it e.g.}, \cite{OrdonezRubin}) that such a contribution is negligible in the case of large $m$ considered here.

\subsection*{Gilkey-de Witt Coefficients}

What is perhaps unusual about the renormalizations required to absorb the UV divergences of $\Vone$ is that they are not done using the couplings of effective interactions in the 4D theory. Because the wavelengths of interest are much shorter than the extra-dimensional size, divergences are instead absorbed into counter-terms in both the 6D bulk and 4D brane actions. Ref.~\cite{Companion} shows how to disentangle which bulk and brane interactions absorb the divergences found in eq.~\pref{eq:Vinfty}. 

\medskip\noindent{\em Bulk divergences}

For our purposes, it is sufficient to notice that the Gilkey coefficients $s_i$ decompose into a bulk and brane part as follows \cite{Companion}:
\be \label{sdecoup}
s_i = f(\alpha,1) \, s_i^{\rm sph} + \delta s_i(\alpha,\Phi) \,,
\ee
where the multiplying factor $f(\alpha,1)$ reflects the change in volume due to the presence of the brane source\footnote{For completeness, $f$ is
\be \label{falphapmdef}
f(\alpha_+,\alpha_-) = \frac{2\,(\alpha_+\alpha_-)^{3/4}}{\sqrt{\alpha_+} + \sqrt{\alpha_-}} \,.
\ee
}. The specific form of $f$ is not needed since the bulk contributions to the Gilkey coefficients are independent of the boundary conditions, and so are guaranteed to cancel---when summed over a multiplet---as they do in the Salam-Sezgin case \cite{KandM,RicciFlatUV}.
%
Physically, this is because the bulk counterterms capture the effects of very short-wavelength modes, which don't extend far enough through the extra dimensions to `know' about conditions imposed at the boundaries. 

\medskip\noindent{\it Brane divergences}

As a corollary to the previous argument: since the brane corrections $\delta s_i$ are capturing the effects of short-wavelength modes at the brane, they depend only on the local properties of that brane, and are insensitive to other non-local, bulk physics. Therefore, there is no need to re-derive these Gilkey-de Witt coefficients; the form which they take in \cite{Companion} is valid for the teardrop as well, despite their original derivation using the rugby ball geometry.

\medskip

In the next section, we will review the cause for interest in a massive multiplet loop and compute its contribution to the renormalized vacuum energy.

\section{Massive Multiplet Loop}
\label{sec:MassMult}

Massive multiplets are of interest in this theory primarily because of the way in which back-reaction determines the four-dimensional vacuum energy. In \cite{TechNat,Companion}, it is shown that the back-reacted, renomalized four-dimensional vacuum energy is given by
\be
\rho_\ssV^{(\ssB\ssR)} = \frac12 \, \frac{\pd \cL_{+\ssR}}{\pd \phi} +\cV_f \,,
\ee
where $\cL_{+\ssR}$ is the renormalized brane lagrangian density: $S_{+\ssR} = \int\!\exd^4x\sqrt{-g}\,\cL_{+\ssR}$. (As mentioned previously, we assume that the first term above dominates, and that $\cV_f$ can be safely dropped in what follows.) The vacuum energy is zero at the classical level since---in the case of interest to us---there is no hidden $\phi$-dependence in either $T$ or $\cA$. Furthermore, we can see that the divergent contributions from massless bulk loops do not contribute to the back-reacted vacuum energy because such loops do not introduce any new $\phi$-dependence; back-reaction simply cancels out the 1-loop contribution along with the classical one.

However, in the case of loops of massive particles, it is possible to obtain a non-zero contribution through back-reaction. This is because---as a result of the classical scale symmetry---masses take the form 
\be
m^2(\phi) = M^2 e^\phi \,,
\ee
and so loops can grow a dilaton dependence in the brane couplings. More explicitly, ref.~\cite{Companion} shows that---although the coefficient $s_{-1}$ and $s_0$ are zero for the massive multiplet---the coefficients $s_1$ and $s_2$ are non-zero, and the renormalized 1-loop potential $V_\ssR = - \cL_\ssR$ is given by 
\be
V_\ssR = \frac{\cC}{(4\pi r^2)^2} \,\ln \left(\frac{m_g(\phi)}{m(\phi)}\right) \,,
\ee
then the corresponding vacuum energy is
\bea
\rho_\ssV &=& -\frac{m^2(\phi)}{2(4\pi)^2 r^2} \frac{\pd \cC}{\pd(m^2)} \ln \left(\frac{m_g(\phi)}{m(\phi)}\right) \\
&=& -\frac{1}{2(4\pi r^2)^2} \bigg[\delta s_{-1} \left(\frac{\kappa M}{2\gR}\right)^6 - \delta s_0 \left(\frac{\kappa M}{2\gR}\right)^4 \nn\\
&&\qquad\qquad\qquad+ \delta s_1 \left(\frac{\kappa M}{2\gR}\right)^2 \bigg] \ln \left(\frac{M_g}{M}\right)  \label{rhovacresult}\,.
\eea
(Recall that we need only the brane divergences since the bulk divergences cancel automatically, as discussed in the previous section.) In the above, the logarithm remains undifferentiated because $m_g$ and $m$ will have identical dependence on $\phi$, as a result of the scaling symmetry. In the second equality, we substitute for $m$, $\cC$ and evaluate using the definition of $r$, namely $r:= \tfrac{\kappa}{2\gR} \,e^{-\phi/2}$.

Given this, it remains only to obtain the Gilkey-de Witt coefficients. Rather than do so again here, we refer the reader to the results in \cite{Companion,LittlestGravitino}; the coefficients are
\be
\delta s_{-1} = \delta s_0 = 0 \,,\quad \delta s_1 = \frac{1-\alpha^2}{2\,\alpha^3} \left(\Phi-\Phi_{\rm s}\right)^2 \,.
\ee
These apply in the regime of interest, where $\Phi >\Phi_{\rm s}:=\pm(1-\alpha)/2$ (as derived in  eqs.~\pref{Phidef}, \pref{ephipval}, \pref{ephisval} from section \ref{sec:Teardrop}). Immediately, we see from eq.~\pref{rhovacresult} that a negative vacuum energy arises, since $\delta s_1 >0$ for any $\alpha <1$.

 Finally, if we substitute the value
\be
\Phi = \pm \big(1-\sqrt{\alpha}\big)
\ee
(as enforced by flux quantization), we find that the renormalized, back-reacted, 1-loop vacuum energy is given by
\be
\rho_\ssV = -\frac{(1-\alpha^2)(1-\sqrt{\alpha})^4}{16\,\alpha^3(4\pi r^2)^2} \left(\frac{\kappa M}{2 \gR}\right)^2 \ln \left(\frac{M_g}{M}\right) \,.
\ee
Keeping in mind the considerations of Section \ref{sec:Teardrop}, we can rewrite this in terms of the brane tension since
\be
\sqrt{\alpha} = 2\,\sqrt{1-\frac{\kappa^2 T}{8\pi}}-1 \simeq 1- \frac{\kappa^2 T}{8\pi} \,.
\ee
We find that---to leading order in $\kappa^2 T$---$\rho_\ssV$ is 
\be
\rho_\ssV \simeq -\frac{1}{4(4\pi r^2)^2} \left(\frac{\kappa^2 T}{8\pi}\right)^5 \left(\frac{\kappa M}{2 \gR}\right)^2 \ln \left(\frac{M_g}{M}\right) \,.
\ee
Many of the bulk constants are tied to the gravity scale, $M_g$; in what follows, let's take
\be
\kappa = M_g^{-2} \,,\quad \gR = (0.001\tilde g) \, M_g^{-1} \,,\quad M = 0.2 \, M_g 
\ee
which together imply
\be
\left(\frac{\kappa M}{2 \gR}\right)^2 =(m r)^2 = \frac{10^4}{\tilde g^2} \,.
\ee
It follows that $(mr)^2 \gg 1$ whenever $\tilde g \lsim 1$, which was assumed earlier in our discussion of the reliability of this result. Substituting, we find
\bea
\rho_\ssV &\simeq& -\rho_\ssV^{\rm obs.} \times \left(\frac{T}{(10 \,{\rm TeV})^4}\right)^5 \left(\frac{(36.3 \,{\rm PeV})^4}{\sqrt{\tilde g}\, r M_g^5}\right)^4 
\eea
(with $\rho_\ssV^{\rm obs.} := (2.3 \times 10^{-3} \,{\rm eV})^4$). Since the radius and the gravity scale are related by
\be
\frac1{\kappa_4^2} \simeq \frac{4\pi r^2}{\kappa^2} = (2.4 \times 10^{18} \,{\rm GeV})^2 \,,
\ee
we are now in a position to state bounds on $M_g$ and $r$ explicitly, which we present in the concluding section.

\section{Conclusion}
\label{sec:Conclusion}

In this paper, we have laid out the conditions under which a massive bulk loop is the dominant contribution to the 1-loop vacuum energy. Although this model may serve well as a benchmark in the space of possible supersymmetric codimension-2 models, it fails to predict the correct sign for the observed vacuum energy. Nevertheless, our order-of-magnitude analysis bodes well for the class of theories to which this one belongs. We obtain the following estimates for gravity scale, $M_g$, and the size of the extra dimensions, $r$:
\bea
M_g &=& \frac{(13.6 \, {\rm TeV})}{\tilde g^{1/6}} \left(\frac{T}{(10\,{\rm TeV})^4}\right)^{5/3} \left(\frac{\rho_\ssV^{\rm obs.}}{|\rho_\ssV|}\right)^{1/12} \\
r &=& (0.73 \, \mu{\rm m})\,\tilde g^{1/3} \left(\frac{(10\,{\rm TeV})^4}{T}\right)^{10/3} \left(\frac{|\rho_\ssV|}{\rho_\ssV^{\rm obs.}}\right)^{1/6} \!\!.\quad
\eea
These values are in agreement with known astrophysical \cite{Raffelt} and inverse-square-law \cite{Adelberger} bounds for tensions near $10$ TeV, but smaller values begin to invalidate the assumption of a perturbatively small $\kappa^2 T$. An interesting feature of these expressions is that the gravity scale and radius satisfy the observational bounds more comfortably as the gauge coupling is decreased.

One caveat to the conclusions drawn here arises in the absence of heavy massive multiplets; in this regime, the finite part of the vacuum energy would dominate. Its precise value (or sign) ought to be a subject for future work, although---in analogy with other known computations of the Casimir energy---it's expected to have a size set by the radius of the extra dimensions, and to vanish in the supersymmetric limit. 

Work is ongoing in developing a four-dimensional effective field theory which captures the extra-dimensional physics at play in these types of brane-world models. The supersymmetric teardrop provides another concrete data point against which to compare such a construction.

\subsection*{Acknowledgments}

Many thanks to Cliff Burgess and Leo van Nierop for helpful discussions. This research has been supported in part by the Natural Sciences and Engineering Research Council of Canada. Research at the Perimeter Institute is supported in part by the Government of Canada through Industry Canada, and by the Province of Ontario through the Ministry of Research and Information.

\begin{appendix}

\section{Details of the Bulk Solution}
\label{appA}

In a warped geometry, it is helpful to use the metric ansatz
\be
\exd s^2 = W^2(\theta) \, \exd s_4^2 + r^2(\theta) \Big(\exd\theta^2 + \alpha^2(\theta) \sin^2\theta \,\exd\varphi^2 \Big) \,.
\ee
(To see a conversion between the polar coordinate $\theta$ and the `exploded' coordinate $\eta$---which is more standard in the literature---see \cite{LittlestGravitino}.) In these coordinates, conical singularities are generally found at $\theta\to\theta_b$ where $\cos\theta_b = b$ and $b:=\pm1$ distinguishes between the north/south brane. The equations of motion yield the following solution \cite{GGP,OtherConical,TechNat}:
\bea
W^4(\theta) &=& \cosh\xi - \sinh\xi \cos\theta \\
r(\theta) &=& r_0 W(\theta) \,,\quad \alpha(\theta) = \frac\lambda{W^4(\theta)}\\
F_{\theta\varphi} &=& \pm \frac{\lambda}{2} \frac{\sin\theta}{W^8(\theta)} = \pm \frac{1}{2\,r^2(\theta)}\frac{\epsilon_{\theta\varphi}}{W^4(\theta)} \\
e^{\phi(\theta)} &=& \frac{e^{\phi_0}}{W^2(\theta)} \,.
\eea
The free parameters are $\lambda$, $\xi$, and $\phi_0$. (In the above, $r_0$ is fixed in terms of $\phi_0$: $r_0 := \tfrac\kappa{2\gR} e^{-\phi_0/2}$.) If $\xi=0$ then there is no warping and this solution reduces to the rugby-ball case. 

Near the poles, we find that the geometry takes the form 
\be
\exd s^2 \Big|_{\theta\to \theta_b} \simeq W_b^2\, \exd s_4^2 + r_b^2 \Big(\exd \theta^2 + \alpha_b^2 (\theta-\theta_b)^2 \exd\varphi^2 \Big)
\ee
where $W_b^4 := e^{-b\xi}$, $r_b := r_0 W_b$, and 
\be \label{alphadef}
\alpha_b := \frac{\lambda}{W_b^4} = \lambda \, e^{b\xi} \,.
\ee
(Conical defects arise whenever $\alpha_b \neq 1$.) Given the absence of a conical defect as $\theta\to\pi$, we find
\be \label{alphadef}
\lambda = e^{\xi} =\sqrt{\alpha_+} \,.
\ee
The other quantities of interest in the main text are the volume of the extra dimensions, and the integral of the flux over the extra dimensions. For completeness, we derive these quantities here from the previously-stated solution.

\medskip\noindent{\em Integral of the Flux}

\medskip\noindent
The flux integral---which appears in the main text in the flux quantization condition---is obtained by computing 
\ba
\!I_\ssF := \int \!\exd^2y \,F_{\theta\varphi} = \pm\pi \lambda \int_{0}^\pi  \!\exd\theta \, \frac{\sin\theta}{W^8(\theta)} \,.
\ea
Evaluating this integral gives
\be
I_\ssF = \pm 2\pi \lambda = \pm 2\pi\sqrt{\alpha_+} 
\ee
which agrees with the form used in the main text.

\medskip\noindent{\em Extra-dimensional Volume}

\medskip\noindent
The volume of the extra dimensions is found by integrating $\exd V_2:= \sqrt{g_2} \, \exd^2y$ over the extra dimensions:
\ba
I_{\ssV_2}:= \int \! \exd^2 y \,\sqrt{g_2} = 2\pi\lambda \, r_0^2  \int_{0}^\pi \!\exd\theta \, \frac{\sin\theta}{W^2(\theta)} \,.\quad\quad
\ea
Evaluating this integral gives
\be
I_{\ssV_2} = \frac{4\pi \lambda\,r_0^2}{\cosh(\xi/2)}
\ee
which, when rewritten using eqs.~\pref{alphadef} and \pref{falphapmdef}, becomes
\be
I_{\ssV_2} = 4\pi r_0^2 \left(\frac{2\,(\alpha_+\alpha_-)^{3/4}}{\sqrt{\alpha_+}+\sqrt{\alpha_-}} \right) = 4\pi r_0^2 \, f(\alpha_+,\alpha_-) \,.
\ee

\end{appendix}

\end{document}